\documentclass[prl,showpacs,twocolumn]{revtex4}%
\usepackage{amsmath}
\usepackage{graphicx}
\usepackage{amsfonts}
\usepackage{amssymb}%
\usepackage{color}

\begin{document}
\title{Multiplexed Memory-Insensitive Quantum Repeaters}
\date{\today }
\author{O. A. Collins, S. D. Jenkins, A. Kuzmich, and T. A. B. Kennedy}
\affiliation{School of Physics, Georgia Institute of Technology,
Atlanta, Georgia 30332-0430} 

\begin{abstract}Long-distance quantum communication via distant pairs of
entangled quantum bits (qubits) is the first step towards more
secure message transmission and distributed quantum computing. To
date, the most promising proposals require quantum repeaters to
mitigate the exponential decrease in communication rate due to
optical fiber losses. However, these are exquisitely sensitive to
the lifetimes of their memory elements. We propose a multiplexing of
quantum nodes that should enable the construction of quantum
networks that are largely insensitive to the coherence times of the
quantum memory elements.
\end{abstract}
\maketitle

Quantum communication, networking, and computation schemes utilize
entanglement as their essential resource. This entanglement enables
phenomena such as quantum teleportation and perfectly secure quantum
communication \cite{bb84}. The generation of entangled states, and
the distance over which we may physically separate them, determines
the range of quantum communication devices. To overcome the
exponential decay in signal fidelity over the communication length,
Briegel {\it et al.} \cite{briegel} proposed an architecture for
noise-tolerant quantum repeaters, using an entanglement connection
and purification scheme to extend the overall entanglement length
using several pairs of quantum memory elements, each previously
entangled over a shorter fundamental segment length. A promising
approach utilizes atomic ensembles, optical fibers and single photon
detectors \cite{duan,matsukevich,chaneliere}.

The difficulty in implementing a quantum repeater is connected to
short atomic memory coherence times and large optical transmission
loss rates. In this Letter we propose a new entanglement generation
and connection architecture using a real-time reconfiguration of
multiplexed quantum nodes, which improves communication rates
dramatically for short memory times.

A generic quantum repeater consisting of $2^N+1$ distinct nodes is
shown in Fig. 1a. The first step generates entanglement between
adjacent memory elements in successive nodes with probability $P_0$.
An entanglement connection process then extends the entanglement
lengths from  $L_0$ to $2L_0$, using either a parallelized (Fig.
1b), or multiplexed (Fig. 1c) architecture. This entanglement
connection succeeds with probability $P_1$, followed by subsequent
entanglement-length doublings with probabilities $P_2$,...,$P_N$,
until the terminal quantum memory elements, separated by
$L={2^N}L_0$, are entangled.

\begin{figure}
  \includegraphics[width=8.6 cm]{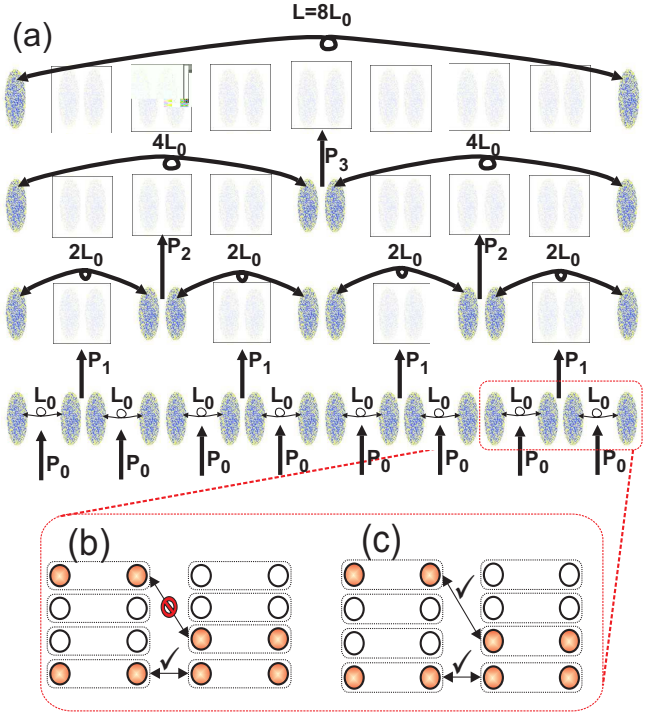}\\
  \caption{(a) Processes of an $N=3$
multiplexed quantum repeater. In addition to two terminal nodes the
network has seven internal nodes consisting of two quantum memory
sites containing $n$ independent memory elements. Entanglement
generation proceeds with probability $P_0$, creating 8 entanglement
lengths of $L_0$. In the lowest panel, shaded memory sites indicate
successfully entangled segments. The $N=1$ level entanglement
connection proceeds with probability $P_1$, producing four entangled
segments of length $2L_0$. Nodes reset to their vacuum states by the
connection are blank. The $N=2$ and $N=3$ levels proceed with
probabilities $P_2$ and $P_3$. Each stage results in
entanglement-length doubling, until an $N=3$ success entangles the
terminal nodes. (b) and (c) show the topology of the $n$ memory
element sets. The parallel architecture (b) connects entanglement
only between memory elements with the same address. In contrast,
multiplexing (c) uses a fast sequential scanning of all memory
element addresses to connect any available memory
elements.}\label{Figure 1}
\end{figure}

For the simplest case of entanglement-length doubling with a single
memory element per site ($N=n=1$), we calculate the average time to
successful entanglement connection for both ideal (infinite) and
finite quantum memory lifetimes. This basic process is fundamental
to the more complex $N$-level quantum repeaters as an $N$-level
quantum repeater is the entanglement-length doubling of two
$(N-1)$-level systems.

{\it{Entanglement-length doubling with ideal memory elements}}.---
Define a random variable $Z$ as the waiting time for an entanglement
connection attempt (all times are measured in units of $L_0/c$,
where the speed of light $c$ includes any material refractive
index). Let $Y\equiv 1$ if entanglement connection succeeds and zero
otherwise. Entanglement generation attempts take one time unit. The
time to success is the sum of the waiting time between connection
attempts and the 1 time unit of classical information transfer
during each connection attempt,

\begin{eqnarray}
&T &=(Z_1+1)Y_1 + (Z_1+Z_2+2)(1-Y_1)Y_2 + \nonumber\\
&&(Z_1+Z_2+Z_3+3)(1-Y_1)(1-Y_2)Y_3 + ...,
\end{eqnarray}
as Z,Y are independent random variables, it follows that
\begin{equation}
\langle T\rangle =\frac{\langle Z\rangle + 1}{P_1},
\end{equation}

In the infinite memory time limit, $Z$ is simply the waiting time
until entanglement is present in both segments, i.e.,
$Z=\max\{A,B\}$, where $A$ and $B$ are random variables representing
the entanglement generation waiting times in the left and right
segments. As each trial is independent from prior trials, $A$ and
$B$ are geometrically distributed with success probability $P_0$.
The mean of a geometric random variable with success probability $p$
is $1/p$, and the minimum of $j$ identical geometric random
variables is itself geometric with success probability $1-(1-p)^j$.
From these properties it follows that,
\begin{equation}
\langle T\rangle _{\infty} =\frac{3-P_0^2}{P_0P_1(2-P_0)}.
\end{equation}

{\it{Entanglement-length doubling with finite memory elements}}.---
For finite quantum memory elements Eqs. (1) and (2) still hold, but
$Z$ is no longer simply $\max\{A,B\}$. Rather it is the time until
both segments are entangled within $\tau$ time units of each other,
where $\tau$ is the memory lifetime. For simplicity, we assume
entanglement is unaffected for $\tau$, and destroyed thereafter. A
new r.v. $M\equiv 1$ if $|A-B| < \tau$, zero otherwise. Due to the
memoryless nature of the geometric distribution,
\begin{eqnarray}
Z &=& \max\{A_1,B_1\} M_1 + \left( \min\{A_1,B_1\}+\tau \right.
\nonumber
\\ &+& \max\{A_2,B_2\})(1-M_1)M_2 + ...
\end{eqnarray}
From this and Eq. (2) it follows that \cite{note}
\begin{eqnarray}
&\langle Z\rangle_\tau &=\frac{1}{P_0(2-P_0-2q_0^{\tau+1})}+
\frac{2\tau
q_0^{\tau+1}}{2-P_0-2q_0^{\tau+1}} \nonumber\\
&& +\frac{2q_0(1-q_0^\tau(1+\tau P_0))}{P_0(2-P_0-2q_0^{\tau+1})},
\nonumber\\
&\langle T\rangle_\tau&=\frac{\langle T\rangle _{\infty}-
(\frac{1+P_0}{P_0P_1})\frac{q_0^{\tau+1}}{1-P_0/2}}{1-\frac{q_0^{\tau+1}}{1-P_0/2}},
\end{eqnarray}
where $q_0 \equiv 1-P_0$. Typically $P_0$ is small compared to $P_1$
as the former includes transmission losses. The terms in $\langle
Z\rangle_\tau$ are, respectively, the time spent(I)waiting for
entanglement in either segment starting from unexcited nodes,(II)
fruitlessly attempting entanglement generation until the quantum
memory in the first segment expires,(III) on successful entanglement
generation in the other segment. When $P_0 \ll 1/(\tau +1)$, memory
times are much smaller than the entanglement generation time, and
$\langle Z\rangle_\tau \approx 1/[P_0^2(1+2\tau)] +
2\tau/[P_0(1+2\tau)] +
 2 \tau^2(1-P_0)/(1+2\tau)$, and term (I) dominates the entanglement-length
doubling time.  Fig. 2 shows the sharp increase in $\langle
T\rangle_\tau$ for small $\tau$ characteristic of term (I).

\begin{figure}
  \includegraphics[width=8.6 cm]{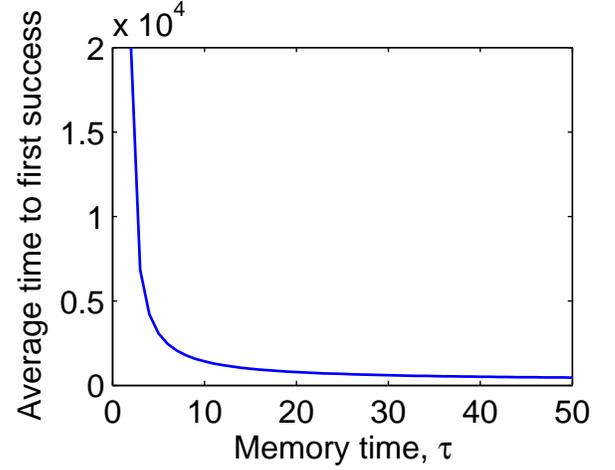}
  \vspace{-0.0cm}
  \caption{Plot of $\langle T\rangle_\tau$ against $\tau$. $P_0=0.01$,
  $P_1=0.5$. The minimum possible success time of 2 imposes a similar
  minimum of the on quantum memory element lifetimes. }\label{Figure 2}
\end{figure}

{\it{Parallelization and multiplexing}}.---  Long memory coherence
times remain an outstanding technical challenge, motivating the
exploration of approaches that mitigate the poor low-memory scaling.
One strategy is to engineer a system that compensates for low
success rates by increasing the number of trials, replacing single
memory elements with $n>1$ element arrays. In a parallel scheme, the
$i^{th}$ memory element pair in one node interacts only with the
$i^{th}$ pair in other nodes, Fig.1b. Thus, a parallel repeater with
$n2^{N+1}$ memory elements acts as $n$ independent $2^{N+1}$-element
repeaters and connects entanglement $n$ times faster. A better
approach is to dynamically reconfigure the connections between
nodes, using information about entanglement successes to determine
which nodes to connect. In this multiplexed scheme the increased
number of node states that allow entanglement connection, compared
to parallelizing, improves the entanglement connection rate between
the terminal nodes.

We now calculate the entanglement connection rate $f_{\tau}$ of an
$N=1$ multiplexed system. Unlike the parallel scheme, however, the
entanglement connection rate is no longer simply $1/\langle
T\rangle_{\tau}$. When one segment has more entangled pairs than its
partner, connection attempts do not reset the repeater to its vacuum
state and there is residual entanglement. Simultaneous successes and
residual entanglement produce average times between successes
smaller than $\langle T\rangle_{\tau}$. We approximate the resulting
repeater rates when residual entanglement is significantly more
probable than simultaneous successes. This is certainly the case in
both the low memory time limit and whenever $nP_0 \ll 1$. Our
approximation modifies Eq. (4) by including cases where the waiting
time is zero due to residual entanglement. In $Z$ of Eq.(4), the
$\min\{ A_i,B_i\}$ terms represent the waiting time to an
entanglement generation success starting from the vacuum state.
Multiplexing modifies Eq.(4) in the following way: for each
$i=1,..\infty$ we replace $(A_i,B_i)$ $\rightarrow$ $
(\min\{A_{i,j}\},\min\{B_{i,k}\})$, where $j$ and $k =1,...,n$. The
effect of the residual entanglement is approximated by the factor
$\alpha$: $\min\{ A_i,B_i\}\rightarrow\alpha \min\{
\min\{A_{i,j}\},\min\{B_{i,k}\}\}$, where 1-$\alpha$ is the
probability of residual entanglement. Eq. (4) now approximates the
average time between successes. Using Eqs. (2), (4) and the
distributions of $\min\{A_{i,j}\}$ and $\min\{B_{i,k}\}$, the
resulting rate is
\begin{eqnarray}
&\langle f\rangle_{\tau,n} &=
\frac{P_1(1-q_0^n)(1+q_0^n-2q_0^{n(\tau+1)})}{1+2q_0^n-q_0^{2n}-4q_0^{n(\tau+1)}+2q_0^{n(\tau+2)}
+
\alpha},\\
&\alpha &= {\frac{q_0^{n-1}(1-q_0^{n})(1-q_0^{2n-1}+2q_0^{3n-2}(1-
q_0^{\tau(2n-1}))}{(1-q_0^{2n-1})(1+q_0^n-2q_0^{(\tau+1)n})}}.
\nonumber
\end{eqnarray}
When $n=1$, $\alpha =1$, as required. Further, as $nP_0,\tau$ become
large, $\alpha \rightarrow 0$ showing the expected breakdown of the
approximation. As $n\rightarrow \infty$, $\alpha$ should approach
$1/2$.

 Fig. 3(a) demonstrates that, as expected,
multiplexed connection rates exceed those of parallelized repeaters.
The improvement from multiplexing in the infinite memory case is
comparatively modest. However, the multiplexed connection rates are
dramatically less sensitive to decreasing memory lifetimes. Note
that the performance of multiplexing $n=5$ exceeds that
parallelizing $n=10$, reflecting a fundamental difference in their
dynamics and scaling behavior. Fig. 3(b) further illustrates the
memory insensitivity of multiplexed repeaters by displaying the
fractional rate $f_\tau/f_\infty$. As parallelized rates scale by
the factor $n$, such repeaters all follow the same curve for any
$n$. By contrast, multiplexed repeaters become less sensitive to
coherence times as $n$ increases. This improved performance in the
low memory limit is a characteristic feature of the multiplexed
architecture.
\begin{figure}
  \includegraphics[width=8.5 cm]{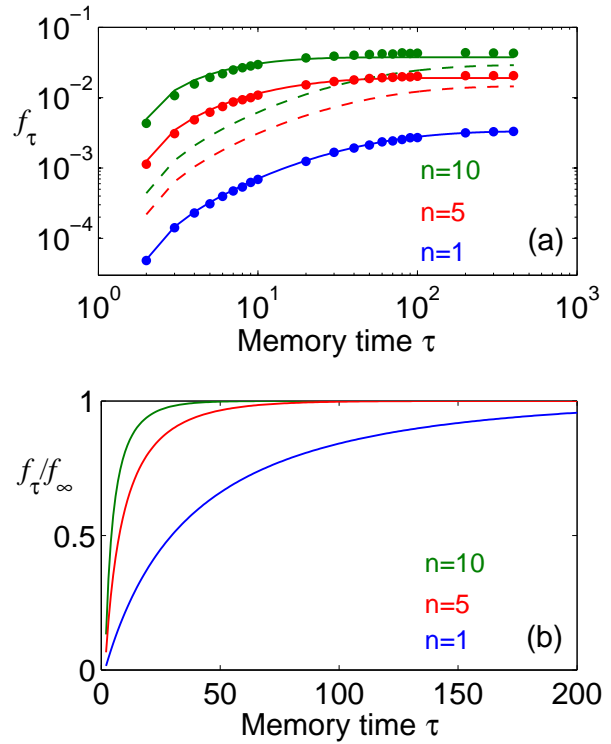}
  \vspace{-0.5cm}
  \caption{Comparison of entanglement connection rate $f_{\tau}$ for
parallel
  (dashed) and multiplexed (solid) architectures as a function of coherence
time. $P_0=0.01$,
  $P_1=0.1$. (a) Solid circles denote simulated values for the multiplexed
  case. (b) Parallel repeaters of any $n$ value follow the $n=1$ line.
  As $n$ increases, multiplexed scaling improves while parallel scaling
remains
  constant.}\label{Figure 3}
\end{figure}

{\it{$N$-level quantum repeaters}}.---  For $N>1$ repeaters we
proceed by direct computer simulation, requiring a specific choice
of entanglement connection probabilities. We choose the
implementation proposed by Duan, Lukin, Cirac, and Zoller (DLCZ)
\cite{duan}. The DLCZ protocol requires a total distance $L$, the
number of segments $2^N$, the loss $\gamma$ of the fiber connection
channels, and the efficiency $\eta$ of retrieving and detecting an
excitation created in the quantum memory elements.

Let $P_0 = \eta_0 \exp(-{\gamma L_0}/2)$, where $\eta_0$ is related
to the fidelity $F\approx 2^N(1-\eta_0)$ \cite{duan}. Recursion
relations give the connection probabilities: $P_i
=({\eta}/(c_{i-1}+1))(1-{\eta}/({2 \beta (c_{i-1}+1)}))$, $c_i
\equiv 2c_{i-1}+1-\eta/\beta$, $i =1,...N$. Neglecting detector dark
counts, $c_0=0$. For photon number resolving detectors $\beta =1$
(PNRDs) \cite{duan}. $\beta =2$ for non-photon resolving detectors
(NPRDs). For values of $\eta < 1$ photon losses result in a vacuum
component of the connected state in either case. For NPRDs, the
indistinguishability of one- and two-photon pulses requires a final
projective measurement, which succeeds with probability $\epsilon =
1/(c_3+1)$, see Ref. \cite{duan} for a detailed discussion.

Consider a 1000 km communication link. Assume a fiber loss of
$10\gamma/\ln10 =0.16$ dB/km, $\eta_0=0.01$, and $\eta=0.9$. Taking
$N=3$ ($L_0=125$ km) gives $P_0=0.001$. For concreteness we treat
the NPRD case, producing connection probabilities: $P_1=0.698$,
$P_2=0.496$, $P_3=0.311$, and $\epsilon=0.206$. Fig. 3 demonstrates
agreement with the exact predictions for $n=1$ and the approximate
predictions for $n>1$. The slight discrepancies for long memory
times with larger $n$ are uniform and understood from the
simultaneous connection successes neglected in Eq. (6).

An $N$-level quantum repeater succeeds in entanglement distribution
when it entangles the terminal nodes with each other. Fig. 4 shows
the entanglement distribution rate of a 1000 km $N=3$ quantum
repeater as a function of the quantum memory lifetime. Remarkably,
for multiplexing with $n\gtrsim10$ the rate is essentially constant
for coherence times over 100 ms, while for the parallel systems it
decreases by two orders of magnitude. For memory coherence times of
less than 250 ms, one achieves higher entanglement distribution
rates by multiplexing ten memory element pairs per segment than
parallelizing 1000. In the extreme limit of minimally-sufficient
memory coherence times set by the light-travel time between nodes,
each step must succeed the first time. The probabilities of
entanglement distribution scale as $nP_0^{2^N}$ (parallelized) and
$(nP_0)^{2^N}$ (multiplexed), for $nP_0 \ll1$.
\begin{figure}
  \includegraphics[width=8.6 cm]{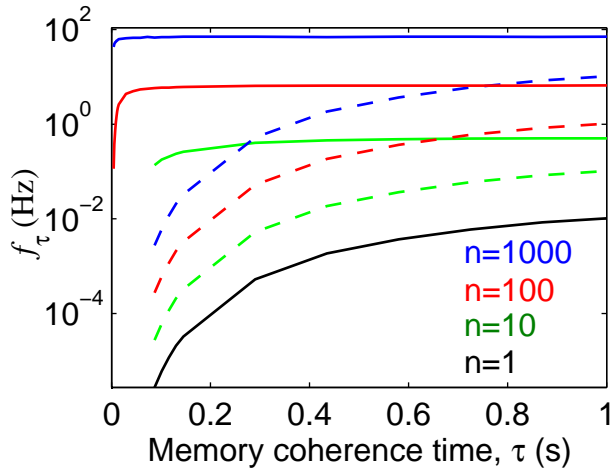}
  \vspace{-0.7cm}
  \caption{Simulated entanglement distribution over $1000$ km for
multiplexed (solid) and
  parallel (dashed) $N=3$ quantum repeaters. For $\tau \geq 100$ msec, the
multiplexed
  distribution rate is almost flat for $n \gtrsim 10$; parallel rates
decrease
  by two orders of magnitude. For low memories a multiplexed $n=10$
  repeater outperforms an $n=1000$ parallelization.}\label{Figure 5}
\end{figure}

{\it{Communication and cryptography rates}}.---  The DLCZ protocol
requires two separate entanglement distributions and two local
measurements to communicate a single quantum bit. The entanglement
coincidence requirement and finite efficiency qubit measurements
result in communication rates less than $f_{\tau}$. Error
correction/purification protocols, via linear-optics-based
techniques, will further reduce the rate and may require somewhat
lower values of $\eta _0$ than the one used in Fig. 4, to maintain
sufficiently high fidelity of the final entangled qubit pair
\cite{duan,bennett1}. We emphasize that it is the greatly enhanced
entanglement distribution rates with multiplexing that make
implementation of such techniques feasible.

{\it{Multiplexing with atomic ensembles}}.---  A multiplexed quantum
repeater could be implemented using cold atomic ensembles as the
quantum memory elements, subdividing the atomic gas into $n$
independent, individually addressable memory elements, Fig. 1c.
Dynamic addressing can be achieved by fast (sub-microsecond),
two-dimensional scanning using acousto-optic modulators, coupling
each memory element to the same single-mode optical fiber. Consider
a cold atomic sample 400 $\mu$m in cross-section in a far-detuned
optical lattice. If the addressing beams have waists of 20 $\mu$m, a
multiplexing of $n>100$ is feasible.  To date, the longest single
photon storage time is 30 $\mu$s, limited by Zeeman energy shifts of
the unpolarized, unconfined atomic ensemble in the residual magnetic
field \cite{dspg}. Using magnetically-insensitive atomic clock
transitions in an optically confined sample, it should be possible
to extend the storage time to tens of milliseconds, which should be
sufficient for multiplexed quantum communication over 1000 km.

{\it{Summary}}.--- Multiplexing offers only marginal advantage over
parallel operation in the long memory time limit. In the opposite,
minimal memory limit, multiplexing is $n^{2^N-1}$ times faster, yet
the rates are practically useless. Crucially, in the intermediate
memory time regime multiplexing produces useful rates when
parallelization cannot. As a consequence, multiplexing translates
each incremental advance in storage times into significant
extensions in the range of quantum communication devices. The
improved scaling outperforms massive parallelization with ideal
detectors, independent of the entanglement generation and connection
protocol used. Ion-, atom-, and quantum dot-based systems should all
benefit from multiplexing.

We are particularly grateful to T. P. Hill for advice on statistical
methods. We thank T. Chaneli\`{e}re and D. N. Matsukevich for
discussions and C. Simon for a communication. This work was
supported by NSF, ONR, NASA, Alfred P. Sloan and Cullen-Peck
Foundations.


\begin{thebibliography}{99}
\bibitem{bb84} C. H. Bennett and G. Brassard, in {\it Proceedings of the
International Conference on Computers, Systems and Signal
Processing} 175 (IEEE, New York, 1984); C. H. Bennett {\it et al.},
Phys. Rev. Lett. {\bf 70}, 1895 (1993); A. K. Ekert, Phys. Rev.
Lett. {\bf 67}, 661 (1991); D. Bouwmeester {\it et al.}, Nature
(London) \textbf{390}, 575 (1997); E. Knill, R. Laflamme, and G. J.
Milburn, Nature (London) \textbf {409}, 46 (2001).
\bibitem{briegel} H. J. Briegel, W. Duer, J. I. Cirac, and P. Zoller, Phys.
Rev. Lett. \textbf{81}, 5932 (1998); W. Duer, H. J. Briegel, J. I.
Cirac, and P. Zoller, Phys. Rev. A \textbf{59}, 169(1999).
\bibitem{duan} L.-M. Duan, M. Lukin, J. I. Cirac, and P. Zoller, Nature
(London) \textbf{414}, 413 (2001).
\bibitem{matsukevich} D. N. Matsukevich and A. Kuzmich, Science {\bf 306},
663 (2004); D. N. Matsukevich {\it et al.}, Phys. Rev. Lett. {\bf
95}, 040405 (2005); D. N. Matsukevich {\it et al.}, Phys. Rev. Lett.
{\bf 96}, 030405 (2006).
\bibitem{chaneliere} T. Chaneli\`{e}re {\it et al.}, Phys. Rev. Lett. {\bf
96}, 093604 (2006).
\bibitem{note} Note that $M$ is independent from $\min\{A,B\}$, but
not from $\max\{A,B\}$, since $\max\{A,B\}\geq |A-B|$. The
calculation of $\langle Z \rangle$ is simplified by using
$\max\{A,B\}=\min\{A,B\}+|A-B|$. Furthermore, as $M$ is either 0 or
1, $\langle M\rangle$ is equal to the probability that $|A-B|<\tau$.
\bibitem{bennett1} C. H. Bennett {\it et al.}, Phys. Rev. Lett. {\bf 76},
722 (1996); S. Bose, V. Vedral, and P. L. Knight, Phys. Rev. A
\textbf{60}, 194 (1999); J.-W. Pan {\it et al.}, Nature (London)
\textbf{410}, 1067 (2001); T. Yamamoto {\it et al.}, Nature (London)
\textbf{423}, 343 (2003).
\bibitem{dspg} D. N. Matsukevich {\it et al.},  Phys. Rev. Lett. {\bf 97},
013601 (2006).
\end{thebibliography}
\end{document}